\documentclass[preprint2]{aastex}
\usepackage{amsmath,amssymb}
\usepackage{graphicx}
\usepackage{rotating}
\usepackage{ulem}
\newcommand{\myemail}{kobelski@solar.physics.montana.edu}

\shorttitle{Multi-Stranded Loops}
\shortauthors{Kobelski et al.}

\begin{document}
\title{Modeling Active Region Transient Brightenings Observed with XRT as Multi-Stranded Loops}
\author{Adam R. Kobelski and David E. McKenzie}
\affil{Department of Physics, PO Box 173840, Montana State University,
	Bozeman, MT, USA, 59717-3840}
	\email{\myemail}
\author{Martin Donachie}
\affil{University of Glasgow, Glasgow, Scotland, UK, G128QQ}
\begin{abstract}
Strong evidence exists that coronal loops as observed in EUV and soft X-rays may not be monolithic isotropic structures, but can often be more accurately modeled as bundles of independent strands. Modeling the observed active region transient brightenings (ARTBs) within this framework allows exploration of the energetic ramifications and characteristics of these stratified structures. Here we present a simple method of detecting and modeling ARTBs observed with the {\it Hinode} X-Ray Telescope (XRT) as groups of 0-dimensional strands, which allows us to probe parameter space to understand better the spatial and temporal dependence of strand heating in impulsively heated loops. This partially automated method can be used to analyze a large number of observations to gain a statistical insight into the parameters of coronal structures including the number of heating events required in a given model to fit the observations. In this article we present the methodology, and demonstrate its use in detecting and modeling ARTBs in a sample data set from {\it Hinode}/XRT. These initial results show that in general, multiple heating events are necessary to reproduce observed ARTBs, but the spatial dependence of these heating events could not yet be established.
\end{abstract}

\keywords{Sun: corona Ñ Sun : flares Ñ Sun: X-rays  Ñ Methods: data analysis}

\section{Introduction}
Flaring structures have been historically represented with models of quickly heated monolithic loops, whose cross-sections can be characterized by a single temperature (\cite{CargillModel1995, Yoshidaetal1995}). This kind of monolithic model has difficulties reproducing the observations, particularly when it comes to reproducing the energy budget and temporal profiles of the temperature, density and emitted radiation (\cite{ReevesWarren}). These monolithic models tend to have plasma cooling faster than observed. One possible explanation for the inconsistency is that the observed single loop is actually a more complex structure that can be described by multiple unresolved strands, with each strand heating and cooling independently \citep[such as shown in:][]{Cargill1994, Warren2006, SarkarWalsh2008}. Recent data from the Hi-C \citep[][]{CirtainHiC2013} sounding rocket strongly support this explanation of threaded loops as do the results of \citet{Brooks2012} using the Extreme Ultraviolet Imaging Spectrometer \citep[EIS;][]{EISInstrument2007}.

Full 3-dimensional modeling of braided strands is computationally intensive, which hinders the models' utility for comparison to observations. When fast computation is required, physical simplifications can be made. The most common simplification derives from the strong magnetic confinement of plasma in coronal strands, which allows the structure to be simplified to one dimension (1D), parallel to the magnetic field. Direct comparison of these 1D models to observations requires the flux tube to be resolved in observation. We can further reduce these 1D models to obtain a zero-dimensional (0D) model. The 0D calculations are commonly performed in two ways. The first method is to simplify the whole strand to the details at a single location along its length (such as done by \citet{CargillModel1995}), the second method is to simplify the equations for parameters averaged over the entire strand (such as done by \citet{EBTEL1}). By utilizing an averaging 0D model to compare to observations, we can achieve quick computations while maintaining an accurate physical description (still bearing in mind the simplifications inherent in collapsing a coronal loop to zero dimensions).

Previous works (e.g. \citet{ReevesWarren, Warren2006}) have modeled the cooling of flares as multi-stranded loops and found that to recreate the observations, multiple strands cooling independently are necessary; but studies like these frequently focus on a single case. \citet{Brooks2012} used {\it Hinode} EIS, and the Atmospheric Imaging Assembly \citep[AIA;][]{AIAInstrument2012} on board the Solar Dynamics Observatory \citep[SDO;][]{SDOInstrument2012} to measure the differential emission measure (DEM) of impulsively heated loops to find lower limits on the number of structures within an active region loop. \citet{Ignacio2013} used AIA to study the temporal profiles of impulsively heated active region loops. To build on these (among other) previous works, we have undertaken a study using active region transient brightenings (ARTBs) observed with the X-Ray Telescope \citep[XRT:][]{Kano2008, Narukage2011, GolubXRT} on {\it{Hinode}} \citep[][]{Kosugi2007} which are modeled with the 0D Enthalpy Based Thermal Evolution of Loops (EBTEL) framework \citep{EBTEL1, EBTEL2, EBTEL3}.

ARTBs are very small flaring events with thermal energy content in the range of $10^{25}-10^{29}$ ergs, compared to the typical $10^{29}-10^{32}$ ergs found in flares. ARTBs occur with a significantly higher frequency \citep[1-40 events per hour per active region -][]{BerghmansMcKenzieClette,Shimizuetal1992}, which provides a large data set for analysis. Here we detect ARTBs in a small (few hours long) data set from XRT, and use models to predict the X-ray flux of the light curves to test the validity of the method.

The detection algorithm is based on the method of \citet{BerghmansClette}, which looks for enhancements in the observed flux on a pixel by pixel basis. Enhancements in neighboring pixels are then grouped together and defined as an ARTB. We use this algorithm on calibrated data from {\it Hinode} XRT of NOAA Active Region 11512 from 22UT on 2012 June 30 through 03:30 UT 2012 July 01. The results of the detection are then background subtracted, and strand lengths are extracted. The detections are then modeled as single- and multi-stranded loops, with each strand modeled using EBTEL. The results of this study show that multiple heating events are needed to reproduce the observed activity, and an impulsive mechanism is required for the triggering of these heating events, which strongly favors reconnection over wave triggering mechanisms.

Section~\ref{artb_meth} describes the detection method, with Section~\ref{artb_par} detailing the parameters of the algorithm. Section~\ref{model} describes the multi-stranded model. In Section~\ref{data} we describe the data, including the details of the background subtraction in Section~\ref{bkgrnd} and the estimation of strand length in Section~\ref{strand_length}. Results are discussed in Section~\ref{results}, with particular details on the multi-stranded cases in Section~\ref{rslt_mult} and the single-stranded cases in \ref{rslt_sing}. We conclude with a discussion in Section~\ref{conc} of the benefits of using the method described here to model and analyze ARTBs from the larger XRT AR catalog.

\section{ARTB Detection}
\subsection{Method Overview}\label{artb_meth}
Initial ARTB detection methods \citep{Shimizuetal1992} relied on visual detection of brightenings in the Geostationary Operational Environmental Satellite (GOES) X-Ray Sensor (XRS) full sun light curves, and {\it Yohkoh}/SXT \citep[][]{SXT1991} image sequences, leaving many smaller but useful observations undetected. Later methods \citep{Shimizu1995} automated the process by analyzing light curves of aggregations of pixels (macropixels) in SXT images looking for sharp increases in flux. \citet{BerghmansClette} developed an improved method relying on a more automated approach to find ARTBs in SXT and EIT \citep[][]{EIT1995}. This approach involves subtracting a running mean from the time profile of brightness in each pixel, then looking for residuals larger than a predefined threshold, typically 3 standard deviations. The automated approach allows greater sensitivity than visual inspection and provides a much more objective process of detection allowing the user to utilize a larger data set. We have adapted and optimized the method of \citet{BerghmansClette} for the data from {\it Hinode}/XRT; herein we present this adapted algorithm and subsequent procedures for fitting models of the coronal structures to the observed light curves.

The first step necessary to utilize the algorithm adapted from \cite{BerghmansClette} is to obtain an appropriate set of calibrated and aligned ARTB observations. After calibration and co-alignment, a running mean of the image stack is created by applying a temporal boxcar smoothing of width $w_{\rm rm}$. This temporally smoothed version of the image is then subtracted from the original image, resulting in a residual with the mean removed.

The next step is to determine the standard deviation of the light curves for each pixel in the image stack, which is used to make a ratio between the residual and the standard deviation. The ``brightest'' pixel in this ratio is determined to be an ARTB as long as it is greater than $q_{\rm D}$, the detection threshold. All pixels in the ratio array that are larger than a second threshold, $q_{\rm C}$, are then checked for connectivity to the $q_D$ point, with all continuously connected points denoted as part of an individual ARTB. All connected points are replaced with the value of the running mean at their location, effectively hiding the detected regions from subsequent iterations of the algorithm.  The detection process then reiterates, ending when the ``brightest'' pixel in the ratio between the running mean subtracted residual and the standard deviation is smaller than the detection threshold.

\subsection{Usage Parameter Overview}\label{artb_par}
Not surprisingly, the efficiency of the ARTB detection algorithm is affected by the values of the constraining parameters $w_{\rm rm}$, $q_{\rm D}$, and $q_{\rm C}$.  In this section, we describe how each of these parameters affects the outcome.

The span used for calculating the running mean, $w_{\rm rm}$, determines in part the sensitivity of the algorithm to secular variations in X-ray flux. This parameter sets the time scale of variations in the running mean which is subsequently removed from the original image sequence. Shorter widths, corresponding to smaller values of $w_{\rm rm}$, result in shorter time scales being removed, and thus smoother residual light curves.  At the extreme, a value of $w_{\rm rm} = 1$ produces a boxcar of unit width; and removal of a running mean so calculated leaves a perfectly smooth (and useless) residual. Expanding the boxcar width by use of larger $w_{\rm rm}$, on the other hand, relaxes the sensitivity of the running mean to short-time-scale fluctuations, so that the running mean (the subtrahend of the residual) becomes more smooth.  Subtraction of a smooth running mean results in a residual that retains much of the sensitivity to transient fluctuations.  The extreme case of this is $w_{\rm rm}$ equal to the number of images in the data stack, such that the subtrahend is tantamount to a simple pedestal, and an equal value is subtracted from every image in the stack.  In such a case, the temporal behavior of the flux is not changed at all.  Thus the optimal value of $w_{\rm rm}$ is that which will accentuate the temporal dynamics of the system, on the time scale of the ARTBs.  We have found, for the test cases considered in this study, $w_{\rm rm}=20$ (images) yields the most consistent results which for this data set corresponds to $\approx10$ minutes.

The unitless threshold for detection, $q_{\rm D}$, has a much more direct effect on the ARTB detection. \citet{BerghmansClette} utilized a threshold of $q_{\rm D}=3$ (i.e., detection threshold of 3 standard deviations above the pre-event mean).  This value, in the test cases we have studied, resulted in an unacceptable number of spurious ``false positives''. This inconsistency with previous results is most likely caused by the increased sensitivity of current solar EUV and X-ray data. Testing with $q_{\rm D}$ = 5 yielded detections of only the largest of ARTBs, or in some cases only flares.  The most reliable results were achieved with $q_{\rm D}$ = 4, although visual confirmation of the resulting light curves is still recommended.

\begin{figure}[ht]
\includegraphics[width=1.\linewidth]{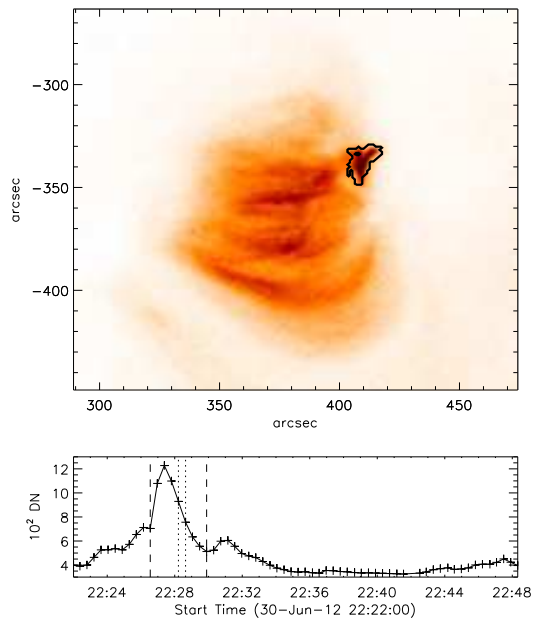}
\caption{Example of a detected ARTB within a subframe of an XRT image. The black contour in the upper reverse color image is the detected region. The lower panel shows the light curve for the detected region over the entire time range for which the algorithm was run. The detection lasts from 02:59:47 to 03:02:42 (dotted lines), and has been manually cropped for analysis from 2:58:32 to 3:06:33 (dashed lines). As mentioned in Section~\ref{results}, the manual cropping is performed to best isolate the complete ARTB light curve from the background.}
\label{examp_lc}
\end{figure}

In \citet{BerghmansClette}, the connectivity threshold $q_{\rm C}$ was held to the same value as $q_{\rm D}$.  Maintaining this equality can be expected to result in underestimating the physical extent of ARTBs where some parts of the heated structure do not reach the same brightness as the peak pixel. The 0D model used here recreates the average parameters of the strand, and so proper measurement of the size of the ARTB better measures the average flux of the observed ARTB, allowing a more accurate comparison between model and observation. In the test cases we considered, as $q_{\rm C}$ was increased, the size of detected ARTBs decreased, eventually becoming single-pixel detections as $q_{\rm C}$ approached $q_{\rm D}$ (with $q_{\rm D}\ge4.0$). Conversely, relaxing $q_{\rm C}$ to lower values can result in many additional pixels being added into the ARTB, even though their fluxes do not increase significantly. When $q_{\rm C}$ was reduced below 3, very large ARTBs were logged, particularly in the final iterations of the scheme (i.e., when many data points had been reset to the median, lowering the required standard deviation threshold).  For most observations $q_{\rm C}$ = 3 was found to be adequate. It is worthwhile to note that at this time only a small number of data sets have been studied, and an expanded search in the future may require the re-evaluation of these parameters. After detection is complete, a 2-D spatial mask of the detected region is created containing every pixel that was flagged by the algorithm.  It is then a trivial task to use this mask to extract light curves for the ARTB detections and normalize the light curves into DN/pixel/sec. An example light curve is shown in Figure~\ref{examp_lc}.  The light curves are then trimmed around the detected ARTB such that the first and last data points are as close to the background level as possible.

\section{Multi-Stranded Model}\label{model}
After detecting ARTBs and extracting their light curves, the next step is to model each ARTB as a loop made of multiple unresolved strands. The modeling process begins by creating a single strand with the 0D EBTEL code developed in \cite{EBTEL1} and revised in \cite{EBTEL3}. EBTEL works by simplifying the one-dimensional time dependent energy equation for a strand into parameters that are spatially averaged across the entire length of the strand. Our observations (as discussed in Section~\ref{artb_meth}) allow this simplification, as we average the flux from the entire observable ARTB. Additionally, works such as \citet{Seaton2001} have shown that ARTB loops appear to brighten in a fairly uniform fashion across the entire loop, though their results focused on EUV observations.

With the minimal input of an assumed heating function and strand length, EBTEL returns the temporal evolution of many important strand parameters, including the spatially averaged temperature, density, and pressure, as well as estimates of the differential emission measure of the chromosphere and corona. EBTEL was chosen for the present study over more complex models for the speed with which it can calculate these parameters. This speed is useful for analyzing many realizations of multi-stranded loops for each ARTB. In seconds, EBTEL is able to accomplish what a multidimensional model might take hours or even days to calculate, with only slight and predictable variations from the results of higher dimensional and more complicated models as discussed in \citet{EBTEL2}.

\begin{figure}[ht]
\includegraphics[width=1.\linewidth]{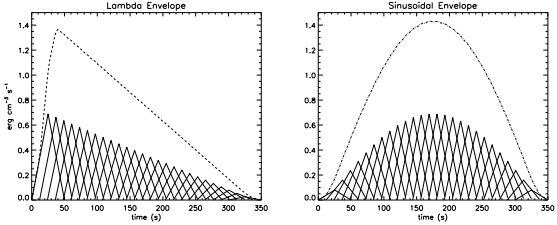}
\caption{An example of the heating function for the case of 26 individual strands with a heating delay of 12 s, heating width of 50s and a peak heating rate of 0.69 ergs cm$^{-3}$ s$^{-1}$. On the left is the lambda shaped envelope, the sinusoidal envelope on the right. The solid lines represent the individual heating events, and the dashed line is the sum of the individual events. These parameters (number of strands/heating events), $\Delta t$, heating width and peak heating rate) are varied for each realization of the model by the genetic algorithm in order to find the best fit between the combined X-ray flux of the strands and flux observed with XRT.}
\label{heatenv}
\end{figure}

One of the primary inputs into the EBTEL model is the heating function. We chose the heating function for each strand to be a triangular shaped function, as can be seen in Figure \ref{heatenv}. This heating function was chosen as a simple proxy for a $\delta$-function, which can be used to create more complicated heating functions, such as when utilizing multiple heating events in the same strand (Section~\ref{rslt_sing}). The other primary input into EBTEL is the half length of the strand, $l_{\rm s}$, which we derive from the observations. It was determined through testing that using EBTEL with timesteps larger than one second when modeling individual strands caused higher peak temperature and density results, while using time steps smaller than 1s did little to change the results from EBTEL. Thus we have chosen a temporal resolution of 1s.

To calculate the column emission measure from the returned density and temperature, we assume each strand to be cylindrical with a constant diameter and axis perpendicular to the line of sight; using these parameters we can convolve the EBTEL results with a given instrument response function (for this study, XRT) to estimate an observable flux for any strand. While the actual strands may not have a constant cross-section, this study looks at the average flux across the entire strand, making the assumption of a uniform cross-section acceptable.

An arbitrary number of unique strands are then created with this method, with each strand heating at a different time. The predicted X-ray fluxes from all of the strands are then superimposed to obtain a model multi-stranded loop. For this proof-of-concept study, we have constrained the independence of individual strands such that the parameters of successive strands are related. The length of all strands in a given ARTB is held fixed, a constraint that is justified by the small size and short duration typical of ARTBs, wherein there is an absence of observational measurements of footpoint spreading, such as flare ribbon or hard X-ray footpoint motions. The delay between successive heating events, $\Delta t$, is also kept the same for all strands, such that strand $n$+1 fires at a set time after strand $n$, constant for all $n$. In the real world, $\Delta t$ is almost certainly more random. It is possible a specific distribution of $\Delta t$ will affect why some loops are brighter/hotter than others, a l\'{a} Parker nanoflares \citep[][]{Parker1988}, but this is a level of complexity to be explored in a future study.

The amplitude of the heating function for each successive strand is varied, though the shape of the individual heating functions remain constant, as does the temporal width of the triangular pulse. For this study, we allow successive strands to heat to peaks dictated by a sinusoidal envelope and a lambda shaped envelope as shown in Figure~\ref{heatenv}. These two envelopes were chosen as they allow us to probe the likelihood of wave heating versus bursty reconnection. It would be expected that wave heating would have a gradual turn-on effect in the heating of individual strands as a particular resonance condition is met \citep[such as with resonant absorption; i.e.][]{Ofmanetal1995}, while bursty reconnection would predict an impulsive beginning to strand heating, followed by a slower decay in the rate of strand energization as system reconfigures. The sinusoidal envelope is symmetric such that the first and last strands have small heating functions and the central strand has the largest heating function, with a sinusoidal shaped envelope between. The lambda shaped envelope has the first strand experiencing the largest heating event, with a linear decline for each successive strand. Each heating envelope is defined by the number of strands ($n$), the delay between heating events ($\Delta t$), the width of the heating pulses ($w$), and the peak heating rate for a strand ($h_{peak}$). A constant (i.e., same for all strands and time invariant) and uniform strand radius ($r_{\rm strand}$) is used to provide an estimate of the volume of plasma emitting in each strand, representing an average radius for all strands in the case that the strands are not uniform.

After superimposing the flux from a given number of strands to create an aggregate light curve, the simulated light curve is compared to the XRT observation by first aligning the time of the peak fluxes. We then increase or decrease the assumed strand radius of the model until the peak fluxes are equivalent. The quality of the fit can then be easily quantified with a simple $\chi^2$ calculation between the observed and modeled flux as shown in the tables below.

To find the best fit for a given observation, the genetic minimization algorithm pikaia is employed \citep[][]{pikaia}. Pikaia is able to adequately and intelligently test parameter space to find the parameters (peak heating rate, width of heat pulses, delay between heating events, number of strands, and strand length) that minimize the $\chi^2$ value between the observed and modeled flux. Pikaia was chosen over other algorithms (such as amoeba or a monte carlo minimization), as it fully samples the parameter space (unlike amoeba) while also intelligently improving its calculation with each iteration (unlike monte carlo method); thus pikaia is fairly resilient to local minima and still relatively quick.

\section{Data}\label{data}
For this study, we use data from {\it Hinode}/XRT of NOAA Active Region 11512 from 22:18UT on 2012 June 30 through 03:35 UT 2012 July 01. These data were split into 4 subsets, separated by observational pauses for environmental conditions (such as the South Atlantic Anomaly) and orbital eclipses. Images were taken in the Ti\_poly filter with a 25s cadence and calibrated using the $\tt{xrt\_prep.pro}$ software suite \citep[][]{xrtprep}. A table of the detections analyzed from this data set can be found in Table~\ref{artb_tab}. Additional ARTBs were detected but not selected for analysis due to complications in their light curves as described in Section~\ref{results}.

\begin{table*}[ht]\scriptsize
\caption{Information on ARTBs detected and analyzed from the XRT data set of AR 11512. The first dataset runs from 22:18UT to 22:48UT on 2012 June 30, the second from 23:25UT to 00:27UT ending on 2012 July 01, the third dataset from 00:58UT to 02:05UT on July 01, and the fourth from 02:37UT to 03:35UT on July 01.}
\begin{tabular}{c|c|c|c|c|c|c|c|c}
\hline\hline
 Data & ARTB & Size of ARTB & Number of & Average & $\chi^2$ (DN) & $\chi^2$ (DN) & $\chi^2$ (DN) & $\chi^2$ (DN) \\
 Set & Number & (pixels/image) & Images & Cadence (s) & Sinusoidal & Lambda & Lambda (Mono) & Single Strand \\
\hline
01 & 000 &  147 &  9 & 22.27 &   22.34 &   11.49 &  776.24 &  874.65\\
01 & 001 &   37 &  9 & 25.79 &   90.78 &    3.82 &    5.68 &   12.41\\
02 & 001 &   52 & 11 & 22.73 &   12.55 &   16.22 &   36.41 &  135.70\\
02 & 002 &   57 &  9 & 22.19 &   27.05 &  648.49 &  196.02 &  359.19\\
02 & 004 &   14 & 11 & 22.64 &   46.57 &  733.45 &   94.66 &  210.79\\
02 & 010 &   21 & 15 & 25.43 &   15.92 &   62.41 &  103.69 &   72.13\\
02 & 014 &   18 & 14 & 23.29 &   97.09 &   51.11 &    2.94 &  504.65\\
02 & 015 &   68 & 20 & 25.18 &  269.59 &   63.72 &   10.62 &  107.94\\
02 & 017 &    9 & 35 & 25.18 & 1682.60 &   33.64 &   25.40 &   87.57\\
02 & 018 &    9 & 20 & 27.73 &   24.05 &   31.32 &   31.75 &   87.69\\
02 & 020 &   27 & 20 & 25.16 &   68.40 &   15.34 &   17.33 &  111.64\\
02 & 021 &   51 & 34 & 25.11 &   26.94 &   26.02 &    6.03 &   67.27\\
02 & 022 &   12 & 17 & 26.46 &  230.15 &   30.83 &    8.54 &   85.92\\
02 & 023 &   39 & 13 & 23.08 &   38.37 &   12.32 &    6.85 &   57.64\\
02 & 024 &    9 & 12 & 22.92 &   60.84 &   13.43 &    6.91 &   27.41\\
02 & 025 &    8 & 19 & 23.67 &   30.20 &   33.48 &   21.43 &   13.16\\
03 & 038 &   23 & 19 & 25.17 &   36.78 &   26.93 &    6.11 &   72.17\\
03 & 040 &   47 & 16 & 23.42 &   41.08 &  172.44 &   13.62 &  301.73\\
04 & 002 &   45 & 16 & 23.42 &   24.32 &   61.26 &   22.84 &   17.54\\
04 & 004 &   99 & 19 & 23.71 &   87.82 &    8.52 &   29.74 &  399.09\\
04 & 006 &   45 & 21 & 23.83 &   30.74 &   37.69 &   16.09 &  103.78\\
04 & 007 &   65 & 19 & 23.71 &   28.64 &   23.30 &   12.60 &  215.38\\
04 & 010 &   28 & 18 & 23.64 &    8.96 &    5.16 &   45.02 &  388.82\\
04 & 011 &    8 & 16 & 23.45 &   35.69 &   25.11 &    4.77 &  613.44\\
04 & 012 &   84 & 19 & 25.30 &   16.16 &  118.33 &   16.15 &    7.58\\
04 & 013 &   10 & 20 & 25.22 &    5.25 &   62.48 &   73.25 &   67.22\\
04 & 014 &   22 & 18 & 25.26 &   20.96 &   26.88 &   19.09 &   19.01\\
04 & 015 &   18 & 19 & 25.30 &   70.99 &   26.01 &   24.55 &   20.69\\
04 & 016 &   48 & 14 & 25.43 &   12.88 &  189.53 &   15.98 &   83.55\\
04 & 022 &   15 & 16 & 23.42 &  122.17 &   83.69 &   43.55 &   83.15\\
04 & 023 &   11 & 12 & 22.92 &  170.25 &   49.75 &   42.78 &   79.27\\
04 & 030 &    9 & 30 & 25.19 &   20.58 &   61.93 &    5.63 &  105.19\\
04 & 031 &    8 & 18 & 25.40 &   38.08 &   42.21 &   63.25 &   81.15\\
04 & 032 &   17 & 19 & 27.64 &   65.66 &  102.10 &   31.16 &  161.86\\
 \hline
\end{tabular}
\label{artb_tab}
\end{table*}

\subsection{Background Subtraction}\label{bkgrnd}
All observations used herein contain background emission from the surrounding corona which will create extra signal on the detector, requiring the model to produce extra flux. It was found in testing that rudimentary background subtraction greatly improves the reliability and consistency of the results. It has been shown in \citet{Klimchuketal1992}, \citet{Schmelzetal2007} and \citet{DeForest1998} that a flat background subtraction can be effective, but can cause ambiguity as to the inferred width of loops. Our background subtraction consists of removing 80\% of the lowest observed flux in the ARTB light curve, where 80\% was chosen to accommodate the flux of the modeled loop existing below the observed noise floor. Since our analysis adjusts the peak flux of the model to match the peak flux of the observation by increasing the radii of the strands, the background flux estimation will have an effect on the model radius. To see this effect, let us first assume the flux of a single strand within one pixel (of width $w_{\rm pixel}$) can be described by
\begin{align}
\Phi_{i} & =g_{i}(T_{i},\rho_{i})\times V_{i} \nonumber \\
 & =g_{i}(T_{i},\rho_{i})\times w_{\rm pixel}\pi r_{i}^2.
\end{align}
where $g_{i}$ is a function similar to the integrated emissivity of the strand at average temperature (T) and density ($\rho$), $V_{i}$ is the volume of a single cylindrical strand of radius $r_{i}$. Thus the total emission of our ARTB will be 
\begin{align}
\Phi_{m0} =\displaystyle\sum\limits_{i=0}^n \Phi_{i}=\displaystyle\sum\limits_{i=0}^n g_i(T_i,\rho_i)\times w_{\rm pixel}\pi r_{i}^2,
\end{align}
Our calculation initially assumes all strands are one pixel wide ($r_{i}=r_{\rm strand}=\frac{w_{\rm pixel}}{2}$) yielding an {\bf unadjusted} model flux $\Phi_{m0}$. $\Phi_{m0}$ is then scaled to match the observed (background subtracted) flux, $\Phi_{\rm obs}$, using the scaling factor $\xi=\Phi_{\rm obs}/\Phi_{m0}$. We arrive at our resultant {\bf adjusted} model flux, $\Phi_m=\xi\Phi_{m0}$. Since we are assuming a constant radius for all strands, $r_{\rm strand}$, $\forall i$
\begin{equation}
\xi=\frac{\Phi_{\rm obs}}{\Phi_{m0}}=\frac{2^2r_{\rm strand}^2}{w_{\rm pixel}^2}.
\end{equation}
Our background subtraction works such that $\Phi_{\rm obs}=\Phi_{\rm data}-\Phi_{\rm bg}$, where $\Phi_{\rm data}$ is the original signal before background subtraction, and $\Phi_{\rm bg}$ is the estimated background signal. Some simple algebra shows that the resultant model strand radius will be:
\begin{align}
r_{\rm strand} & = \frac{w_{\rm pixel}}{2}\left(\frac{\Phi_{\rm data}-\Phi_{\rm bg}}{\Phi_{m0}}\right)^{\frac{1}{2}} \nonumber \\
& =\frac{w_{\rm pixel}}{2}\left(\frac{\Phi_{\rm data}}{\Phi_{m0}}\right)^{\frac{1}{2}}\left(1-\frac{\Phi_{\rm bg}}{\Phi_{\rm data}}\right)^{\frac{1}{2}}.
\end{align}
A brief expansion of the $\Phi_{\rm bg}$ parenthetical shows that 
\begin{align}
r_{\rm strand}\approx&\frac{w_{\rm pixel}}{2}\left(\frac{\Phi_{\rm data}}{\Phi_{m0}}\right)^{\frac{1}{2}}\nonumber\\
&\times\left[1-\frac{1}{2}\frac{\Phi_{\rm bg}}{\Phi_{\rm data}}-\frac{1}{8}\left(\frac{\Phi_{\rm bg}}{\Phi_{\rm data}}\right)^{2}\right.\nonumber\\
&\left.-\frac{3}{48}\left(\frac{\Phi_{\rm bg}}{\Phi_{\rm data}}\right)^{3}+\cdots\right].
\end{align}
This illustrates that while the $\Phi_{\rm bg}$ can have a large effect on the measured radius, small changes in $\Phi_{\rm bg}$ do not. Our background subtraction provides a reasonable estimate of the base level, but small corrections to this level (such as could be found with a more sophisticated method) will not drastically change our results. Hence we determine that our method of background subtraction is sufficient, but that our estimates of strand radius should not be taken too literally.

\subsection{Strand Length}\label{strand_length}
To better constrain the parameter space searched, we have estimated the full length ($L_{\rm obs}$) for each ARTB from images of its detection. To account for 3D projection effects (among other factors), we allow the length to vary, but constrain the parameter search on strand length to stay between $L_{\rm obs}/2.5\le l_{\rm s} \le 2.5L_{\rm obs}$. The lower bound allows the strands to be slightly smaller than the observed length (note that $L_{\rm obs}$ is full length and $l_{\rm s}$ the half length). The upper bound allows the strand (if assumed to follow an elliptical arc) to be 4 times taller than it is wide, and thus most physical strand lengths are contained within our model. The added tolerance for smaller and larger lengths allows a sanity check on our results. 

\section{Results and Analysis}\label{results}
After calibrating the data, the spacecraft pointing jitter was removed and an iterative cross-correlation algorithm was employed individually to each of the 4 datasets to further co-align the data. The datasets were then passed through the ARTB detection algorithm, which resulted in 10 detections in the first dataset (from 22:18UT to 22:48UT), 32 detections in the second (from 23:25UT to 00:27UT), 51 detections in the third (from 00:58UT to 02:05UT), and 33 detections in the fourth (from 02:37UT to 03:35UT). We manually cropped the light curves so that they begin and end as near to the lowest observed flux level as possible, as shown in Figure~\ref{examp_lc}. 

Upon visual inspection of the extracted light curves, only a few of these detections were found to be problematic. One type of problematic detection involves coherent spikes in the quiet regions surrounding an active region. Since these detections are not valid ARTBs and we cannot easily determine whether these signals are quiet sun brightenings or instrument noise, they are not analyzed. Another source of problematic detections are events for which flux enhancements in the light curves overlap in time, creating a more complicated profile. These events have slightly different spatial footprints, but overlapping light curves. This overlap adds a level of complexity to the analysis that we determined was not suitable for a proof-of-concept study such as this, and as such do not analyze these ARTBs here. In addition, some of the detected ARTBs spanned fewer than 7 images, and thus are over-constrained by our model. This concept of over-constraint plays an important role in the simplifications used in our model. Increasing the complexity of the model would increase the number of parameters used by the model and thus the number of time steps required for analysis, further decreasing the number of useable observations. Removing the spurious, over-constrained and multi-peaked detections, we are left with 44 ARTBs for analysis, 34 of which were successfully modeled. Only those that were successfully modeled are shown in Table~\ref{artb_tab}.

We then used pikaia with our multi-stranded model and both (sinusoidal and lambda) heating envelopes to determine the parameters which best fit each ARTB. We allowed the peak heating function to vary from 0.2 to 1.0 ergs cm$^{-3}$ s$^{-1}$, the width of the heating pulse to vary between 2 and 50s (in one second intervals), the delay between successive heating events to vary between 1 and 20s, the number of strands to vary between 1 and 500, and the strand half-length to vary between  $L_{\rm obs}/2.5\le l_{s} \le 2.5L_{\rm obs}$ as discussed in Section~\ref{strand_length}. We have also fit the data with a single strand, both by using a single impulsive heating event and multiple heating events. When considering multiple heating events in a single strand, we have confined our search to using the lambda envelope to constrain the individual heating events, such as represented by the dashed lines in Figure~\ref{heatenv}.

\subsection{Multi-Stranded Results}\label{rslt_mult}
\begin{figure*}[]
\includegraphics[width=.5\linewidth]{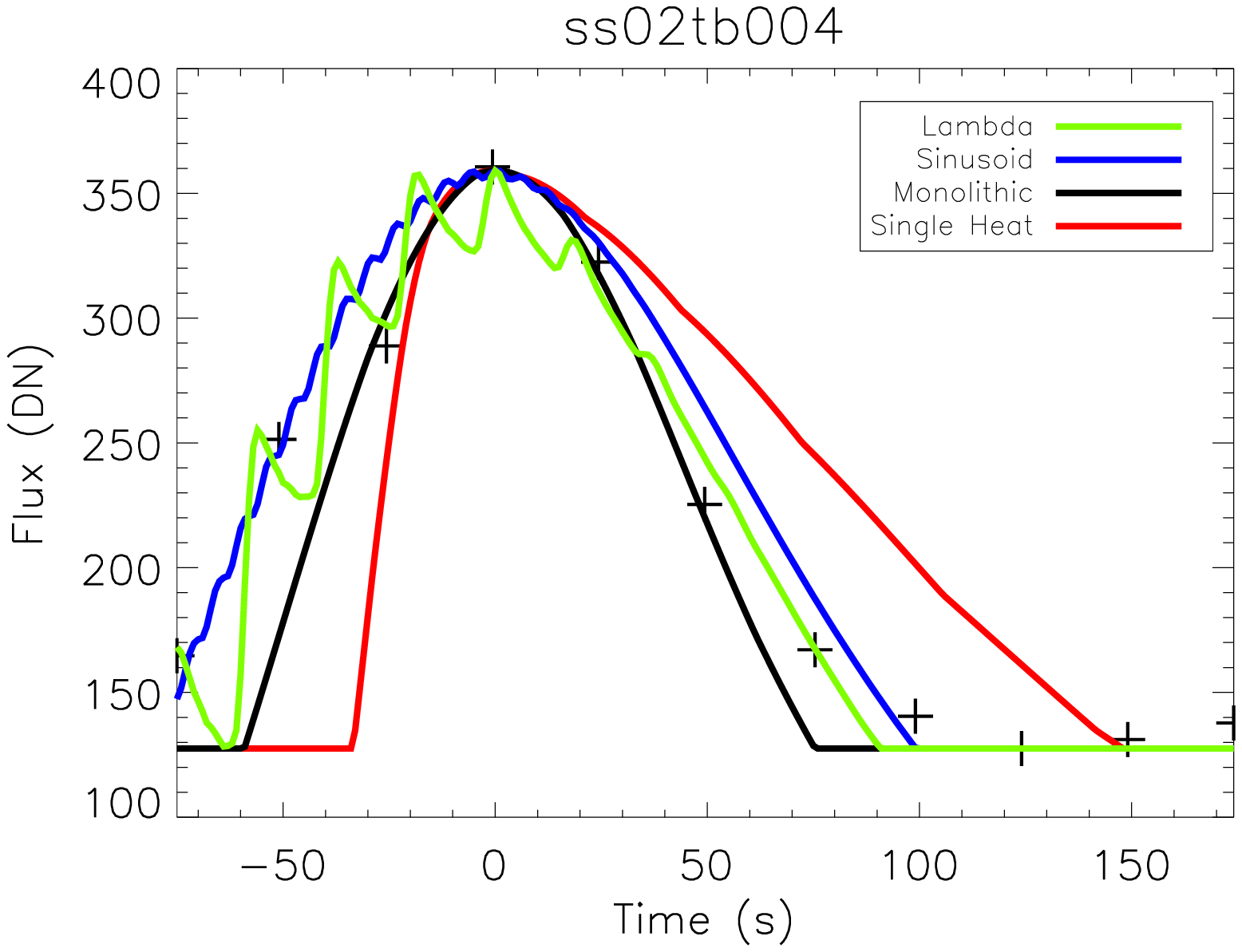}\includegraphics[width=.5\linewidth]{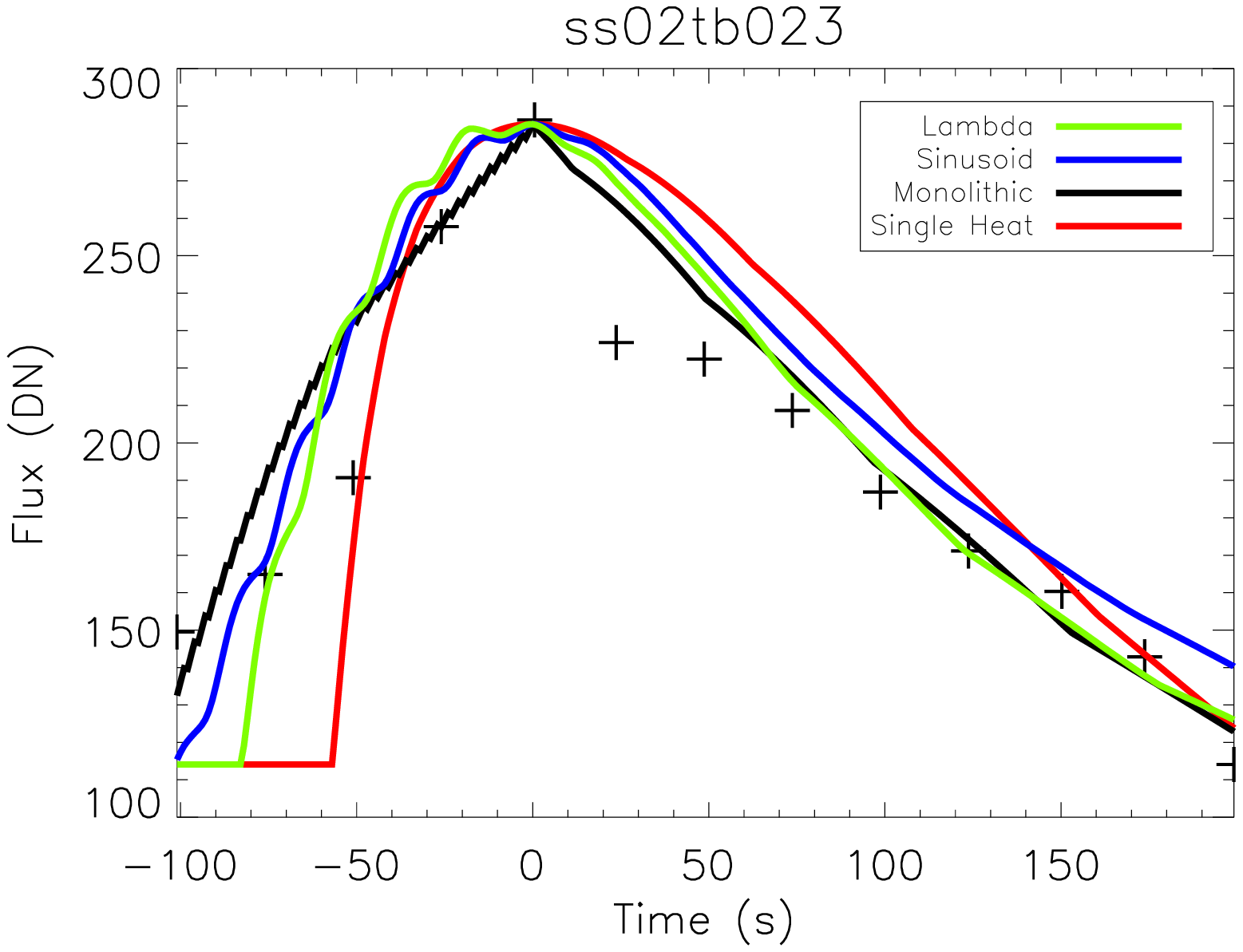}
\includegraphics[width=.5\linewidth]{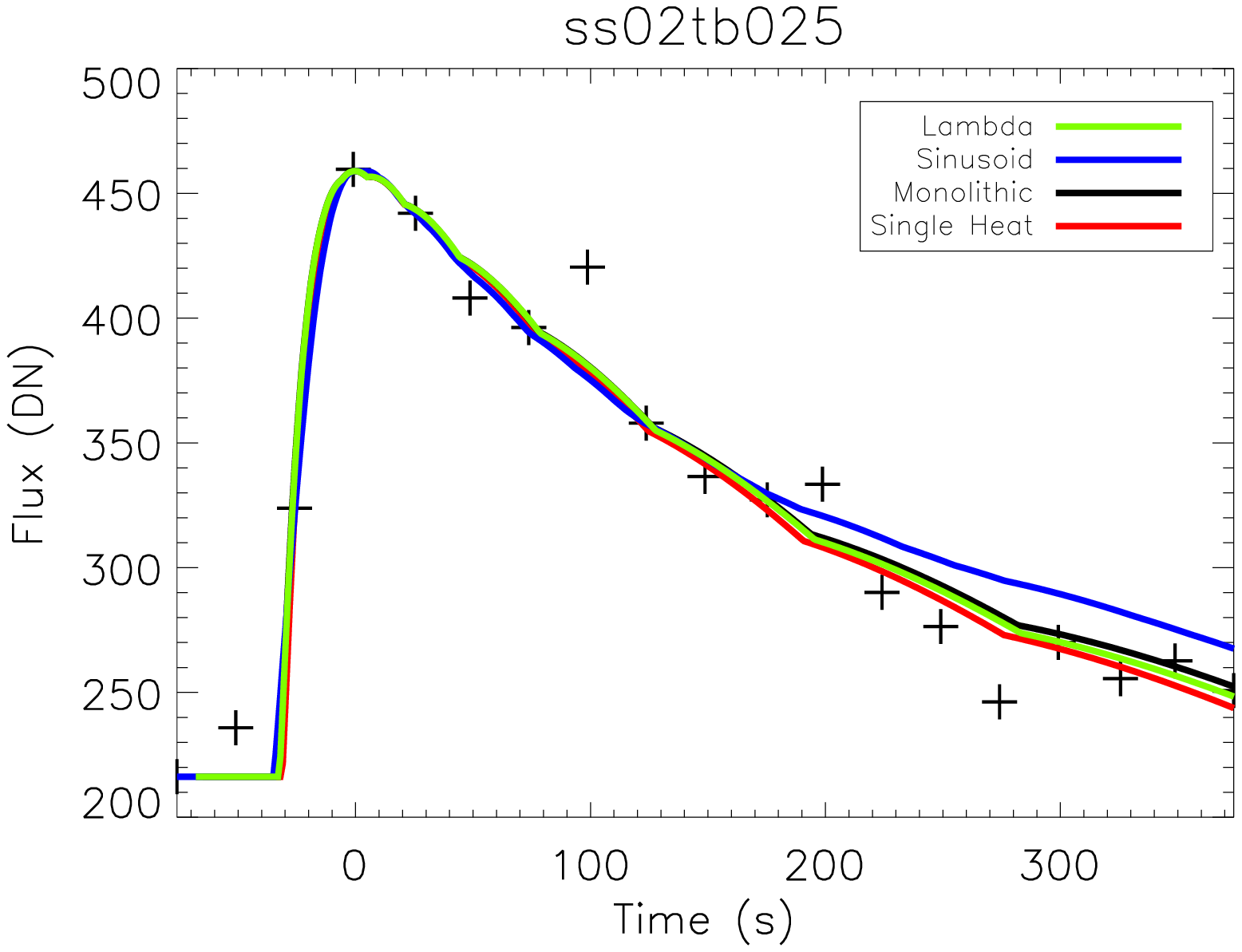}\includegraphics[width=.5\linewidth]{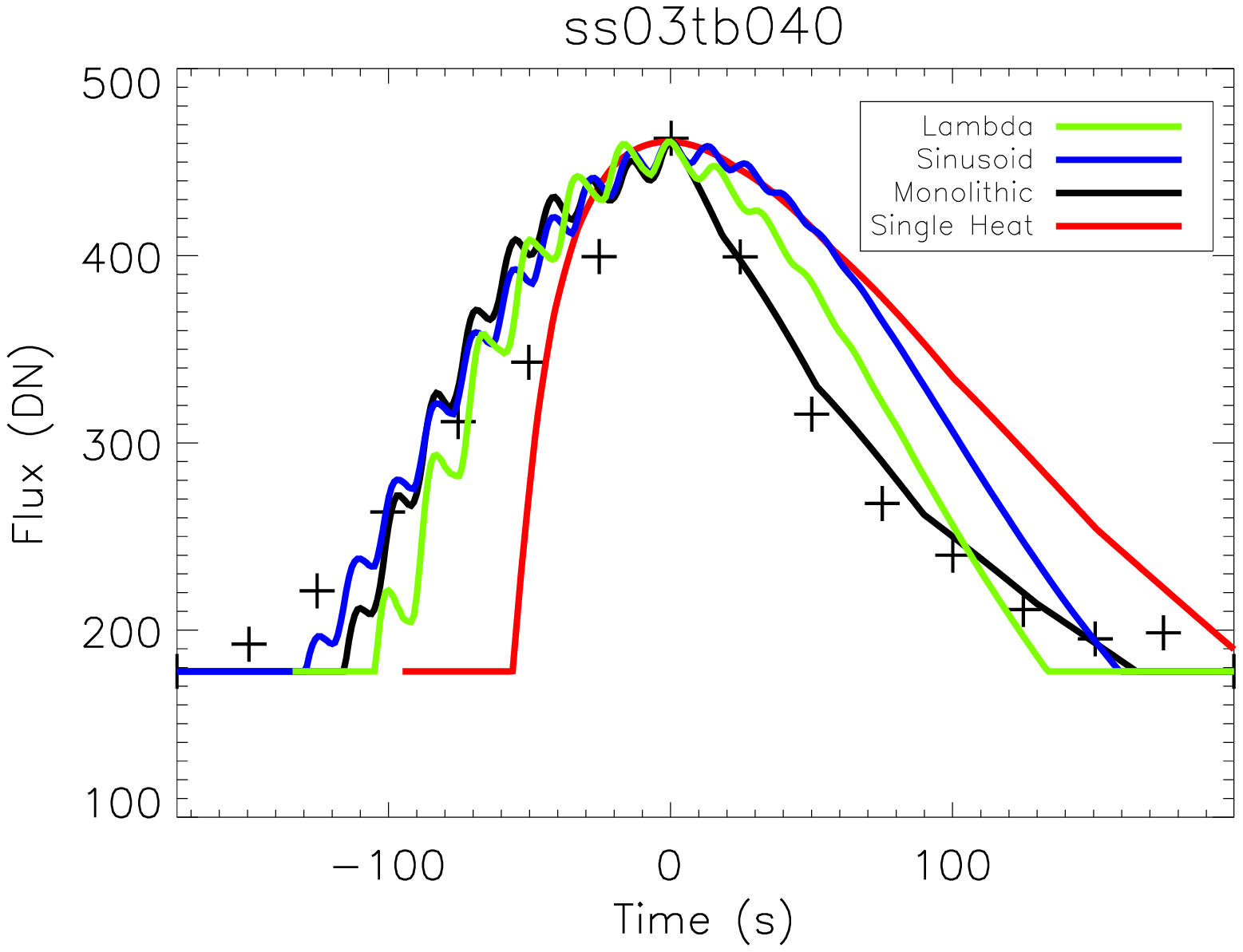}
\caption{Representative fits from 4 ARTBs. Crosses represent the observed light curves after background subtraction. The green and blue lines represent the best fit for the multi-stranded models using the lambda and sinusoidal envelopes respectively. The red line represents the best fit using a single strand heated a single time. The black line represents a single strand heated multiple times using the lambda envelope. The upper left is ARTB 4 from dataset 2; upper right ARTB 23 from dataset 2; lower left ARTB 25 from dataset 2; and the lower right is ARTB 40 from dataset 3. These results show the improvement from using the lambda shaped envelope with either a single strand or a multi-stranded loop. The lower left shows a unique case where a single heating event in one strand can give comparable results to more complex models.}
\label{samp_fit_good}
\end{figure*}

We have modeled 44 detected ARTBs using both the sinusoidal and lambda envelopes. Example fits can be found in the green and blue lines in Figure~\ref{samp_fit_good}. Tables~\ref{sin_res_tab}~and~\ref{lda_res_tab} show the results of the multi-stranded runs. In general (but not in every case), the lambda envelope resulted in lower $\chi^2$ values in the best fits than the sinusoidal envelope. Even though the $\chi^2$ minimization provides a good method for comparing fits when modeling a single ARTB observation, it is limited when comparing results from different observed ARTBs. We have found that comparing the ratio of the integrated flux of the model to that of the observation provides a very good proxy for the quality of the individual results. Visual inspection of the results found that good fits exist when this ratio is between 0.95 and 1.06, but acceptable fits can exist with larger or smaller values. These tests indicated 34 of the 44 ARTBs were acceptably modeled with the both the lambda and sinusoidal envelope. 

When considering those ARTBs that could be fit with these envelopes, the lambda envelope results used a median value of  23 strands with 17s wide heating pulses to fit the observations, while the sinusoidal envelope required 31 strands each being heated for 21.5s (average value for all results). On average, the results for the other parameters were similar for both envelopes. The best fit for the lambda envelope resulted in a single strand only once, for ARTB 25 in dataset 2 (shown in the lower left of Figure~\ref{samp_fit_good}), giving a good visual fit. This illustrates the benefit of such a statistical study: by viewing the results of many runs, it is possible to identify such a result as an outlier, instead of interpreting this as proof of singly heated monolithic loops. On average for the good fits, the $\chi^2$ values for the lambda envelope were 48\% lower than the sinusoidal $\chi^2$, suggesting a significant improvement with the lambda envelope. The suggested strand radius is of order 0.5" for both envelopes, which is slightly larger than expected, though given our rudimentary and minimal background subtraction is not unacceptable. A few results returned unreasonably large strand radii (notably ARTBs 2 and 14 in dataset 2). A possible cause of these large radii is the removal of too small of a background signal. These large radius results accompany low peak heating rates and somewhat large values of their integral ratio suggesting these cases are not well fit by the model and could be readily and consistently flagged.

\begin{figure*}[]
\includegraphics[width=1.\linewidth]{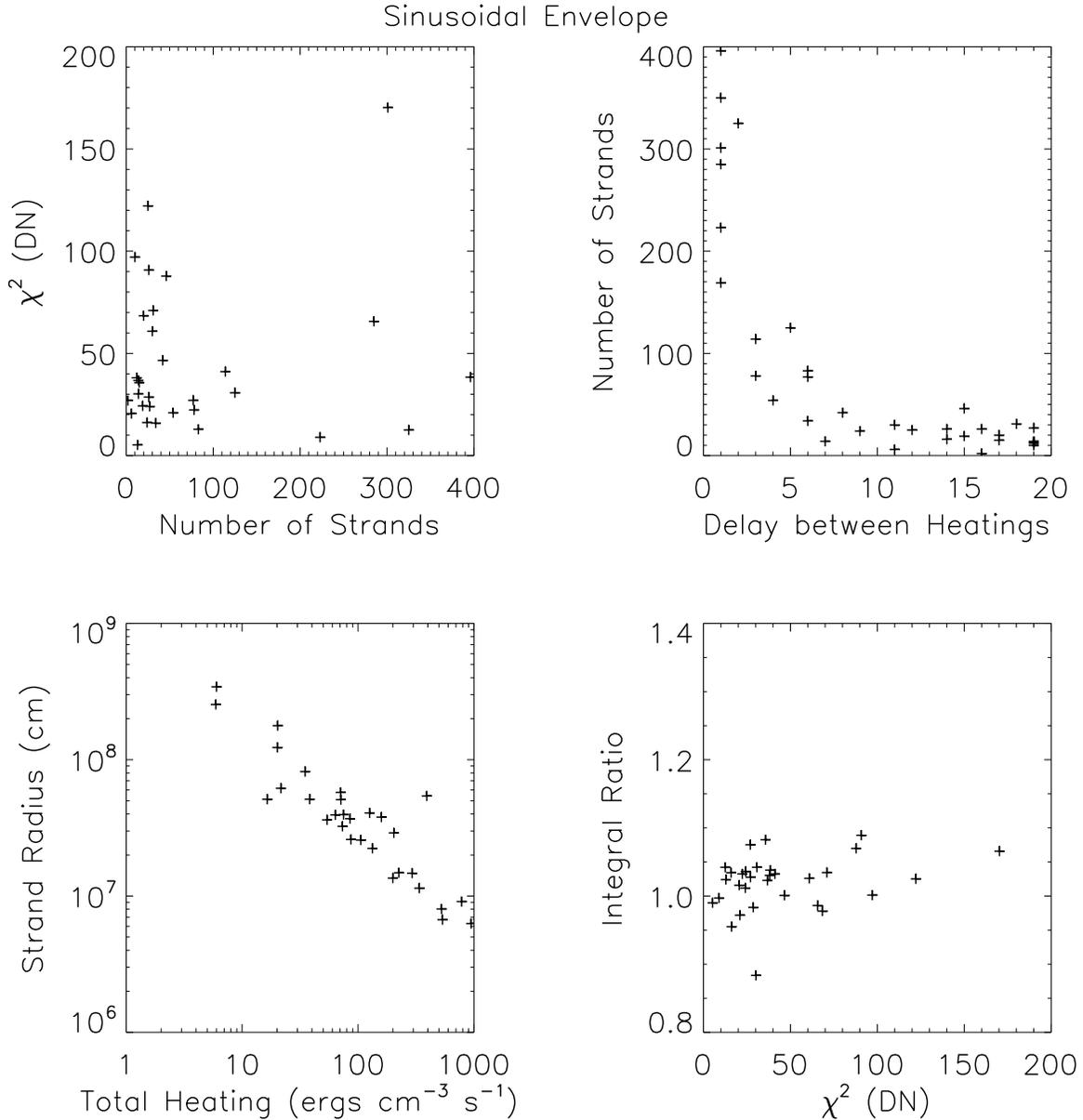}
\caption{Scatter plots for best multi-stranded fits using a sinusoidal envelope: $\chi^2$ versus number of strands ($n$: top left), heating delay ($\Delta t$) versus $n$ (top right), total integrated heating versus strand radius ($r_{\rm strand}$: bottom left), and $\chi^2$ versus ratio of integrated flux (bottom right). For the relation between $\Delta t$ and $n$, the sinusoidal envelope shows a Spearman correlation coefficient of -0.8344 with a chance probability of less than 0.01\%. These comparisons suggest a very strong relation between $\Delta t$ and $n$. The relation between strand radius and input heating is shown. The scatter in the integral ratio plot shows the benefit of using the integral ratio technique to compare results between ARTBs. The other parameter relations (not shown) did not show useful relationships with so few data points.}
\label{res_scat_sin}
\end{figure*}

\begin{figure*}[]
\includegraphics[width=1.\linewidth]{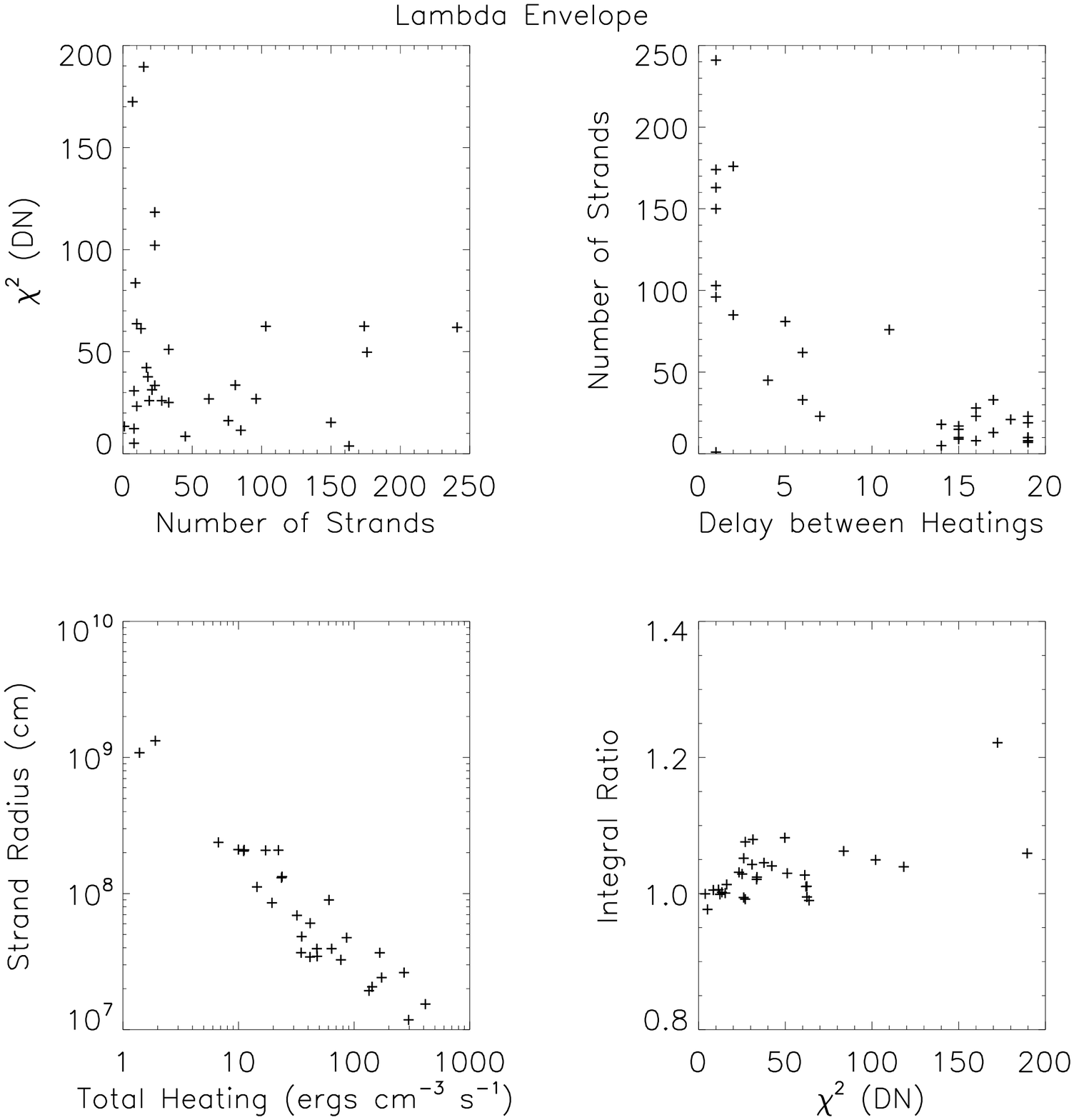}
\caption{Scatter plots for best multi-stranded fits using a lambda envelope, the layout is the same as Figure~\ref{res_scat_sin}. For the relation between $\Delta t$ and $n$, the lambda envelope shows a Spearman correlation coefficient of -0.6736 with a chance probability of less than 0.01\%. These comparisons suggest a very strong relation between the $\Delta t$ and $n$. As should be expected, there is a relation between strand radius and input heating, as more input heating will mean a brighter strand and thus a smaller radius is required. }
\label{res_scat_lda}
\end{figure*}

A desired goal for this and future studies is to search for correlations between the parameters giving the best fits. When looking at the relations between parameters, $\Delta t$ and $n$ have the only notable correlation, as shown in Figures~\ref{res_scat_sin}~and~\ref{res_scat_lda}. A Spearman rank order correlation between $\Delta t$ and $n$ shows correlation coefficients of -0.8344 and -0.6736 for the sinusoidal envelope and lambda envelopes respectively. The chance probabilities for these distributions are less than 0.01\%, suggesting that these parameters are correlated. One should expect some relation between $\Delta t$, $n$, and $w$, as combined they dictate the total width of the heating envelope, though no monotonic relation between these parameters was found. In fact, when looking at the total integrated heat input, no notable correlations were seen, except for a relation with strand radius. This is an expected relation; if more energy is deposited, the strands will be brighter, and thus a smaller strand radius is necessary to match the observations (as suggested by the Equations of Section~\ref{bkgrnd}). Additionally, the median value of 31 strands and 23 strands for the sinusoidal and lambda envelopes (respectively) show a preference for a small number of strands (but generally more than one).

\begin{sidewaystable*}[]
\caption{Best fit results from the using the Sinusoidal Envelope.}
\begin{tabular}{c|c|c|c|c|c|c|c|c|c|c}
\hline\hline

 Data & ARTB & $\chi^2$ & Peak Heating & Heat Pulse & Event & Number of & Half-Length & Half-Length & Strand & Integral\\
 Set & Number & (DN) & (ergs/cm$^{3}$/s) & Width (s) & Delay (s) & Strands &  Observed (Mm) & Result (Mm) & Radius (Mm) & Ratio\\
\hline
01 & 000 & 1682.60 & 0.219 & 16 & 01 & 350 &   8.57 &   2.26 & 0.542 & 1.267\\
01 & 001 &   26.94 & 0.865 & 22 & 16 & 002 &   6.78 &   7.44 & 0.513 & 1.075\\
02 & 001 &   30.20 & 0.316 & 04 & 07 & 014 &   4.91 &   2.44 & 3.428 & 0.884\\
02 & 002 &  230.15 & 0.419 & 47 & 01 & 169 &   6.18 &   5.09 & 0.068 & 1.172\\
02 & 004 &   15.92 & 0.799 & 07 & 06 & 034 &   2.97 &   2.75 & 0.509 & 1.034\\
02 & 010 &   24.05 & 0.659 & 33 & 19 & 027 &   5.37 &  10.59 & 0.135 & 1.012\\
02 & 014 &   97.09 & 0.943 & 26 & 19 & 010 &   4.24 &   3.40 & 0.369 & 1.001\\
02 & 015 &   38.37 & 0.730 & 19 & 01 & 396 &   8.58 &   9.39 & 0.037 & 1.038\\
02 & 017 &  269.59 & 0.622 & 06 & 14 & 016 &   4.03 &  14.74 & 1.230 & 0.946\\
02 & 018 &   60.84 & 0.404 & 15 & 11 & 030 &   5.37 &  16.63 & 0.394 & 1.026\\
02 & 020 &   68.40 & 0.880 & 17 & 17 & 020 &   7.83 &   6.27 & 0.257 & 0.978\\
02 & 021 &   27.05 & 0.357 & 08 & 06 & 077 &  12.20 &  25.92 & 0.576 & 1.028\\
02 & 022 &   30.74 & 0.661 & 05 & 05 & 125 &   6.49 &   7.02 & 0.379 & 1.042\\
02 & 023 &   35.69 & 0.888 & 12 & 17 & 015 &   6.68 &   8.43 & 0.362 & 1.083\\
02 & 024 &   20.96 & 0.653 & 45 & 04 & 054 &   3.59 &   3.31 & 0.080 & 0.972\\
02 & 025 &   20.58 & 0.453 & 06 & 11 & 006 &   6.97 &  17.44 & 2.544 & 1.016\\
03 & 038 &   38.08 & 0.929 & 09 & 19 & 012 &   5.22 &  10.04 & 0.513 & 1.030\\
03 & 040 &   90.78 & 0.591 & 03 & 14 & 026 &   5.30 &   4.63 & 1.778 & 1.089\\
04 & 002 &   16.16 & 0.257 & 35 & 09 & 024 &   6.46 &   9.04 & 0.325 & 0.955\\
04 & 004 &   70.99 & 0.789 & 27 & 18 & 031 &   8.36 &   7.11 & 0.148 & 1.034\\
04 & 006 &  122.17 & 0.710 & 05 & 12 & 025 &   7.54 &   7.32 & 0.818 & 1.025\\
04 & 007 &   28.64 & 0.969 & 15 & 16 & 026 &   9.48 &   7.82 & 0.224 & 0.983\\
04 & 010 &   46.57 & 0.685 & 35 & 08 & 042 &   5.42 &   9.21 & 0.114 & 1.001\\
04 & 011 &  170.25 & 0.958 & 49 & 01 & 301 &   8.46 &   7.52 & 0.026 & 1.066\\
04 & 012 &    5.25 & 0.406 & 12 & 19 & 013 &   8.96 &  20.46 & 0.617 & 0.990\\
04 & 013 &   12.88 & 0.342 & 32 & 06 & 083 &   5.37 &  14.58 & 0.147 & 1.024\\
04 & 014 &    8.96 & 0.443 & 29 & 01 & 223 &   6.65 &  16.27 & 0.063 & 0.997\\
04 & 015 &   22.34 & 0.592 & 35 & 03 & 078 &   4.84 &  16.86 & 0.067 & 1.033\\
04 & 016 &   12.55 & 0.918 & 49 & 02 & 325 &   7.75 &   6.20 & 0.018 & 1.042\\
04 & 022 &   36.78 & 0.761 & 24 & 19 & 014 &   5.34 &   9.28 & 0.260 & 1.023\\
04 & 023 &   24.32 & 0.454 & 25 & 15 & 019 &   3.56 &   5.01 & 0.396 & 1.036\\
04 & 030 &   65.66 & 0.615 & 13 & 01 & 285 &   4.89 &  17.64 & 0.091 & 0.986\\
04 & 031 &   41.08 & 0.348 & 15 & 03 & 114 &   4.47 &  12.52 & 0.291 & 1.033\\
04 & 032 &   87.82 & 0.602 & 13 & 15 & 046 &   7.01 &   7.06 & 0.407 & 1.070\\

\hline
\end{tabular}
\label{sin_res_tab}
\end{sidewaystable*}

\begin{sidewaystable*}[]
\caption{Best fit results from the using the Lambda Envelope.}
\begin{tabular}{c|c|c|c|c|c|c|c|c|c|c}
\hline\hline

 Data & ARTB & $\chi^2$ & Peak Heating & Heat Pulse & Event & Number of & Half-Length & Half-Length & Strand & Integral\\
 Set & Number & (DN) & (ergs/cm$^{3}$/s) & Width (s) & Delay (s) & Strands &  Observed (Mm) & Result (Mm) & Radius (Mm) & Ratio\\
\hline
01 & 000 &  648.49 & 0.997 & 32 & 14 & 005 &   8.57 &   6.86 & 0.346 & 1.327\\
01 & 001 &  733.45 & 0.610 & 03 & 15 & 010 &   6.78 &   2.93 & 2.380 & 0.940\\
02 & 001 &   26.93 & 0.455 & 02 & 01 & 096 &   4.91 &   3.98 & 2.085 & 1.076\\
02 & 002 &  172.44 & 0.332 & 02 & 19 & 007 &   6.18 &   4.96 & 10.832 & 1.222\\
02 & 004 &    5.16 & 0.804 & 08 & 19 & 008 &   2.97 &   2.61 & 1.119 & 0.977\\
02 & 010 &   25.11 & 0.200 & 14 & 17 & 033 &   5.37 &   5.96 & 1.332 & 1.029\\
02 & 014 &   83.69 & 0.365 & 02 & 15 & 009 &   4.24 &   4.79 & 13.274 & 1.062\\
02 & 015 &   49.75 & 0.997 & 49 & 02 & 176 &   8.58 &  10.44 & 0.020 & 1.082\\
02 & 017 &   62.41 & 0.389 & 05 & 01 & 103 &   4.03 &  20.12 & 0.898 & 1.011\\
02 & 018 &   51.11 & 0.490 & 10 & 06 & 033 &   5.37 &  13.30 & 0.606 & 1.030\\
02 & 020 &   63.72 & 0.640 & 20 & 19 & 010 &   7.83 &   6.61 & 0.483 & 0.990\\
02 & 021 &   33.64 & 0.263 & 16 & 05 & 081 &  12.20 &  22.18 & 0.474 & 1.024\\
02 & 022 &   31.32 & 0.880 & 02 & 18 & 021 &   6.49 &  10.10 & 2.105 & 1.080\\
02 & 023 &   30.83 & 0.860 & 17 & 19 & 008 &   6.68 &   8.40 & 0.368 & 1.043\\
02 & 024 &   12.32 & 0.841 & 21 & 16 & 008 &   3.59 &   4.50 & 0.342 & 0.999\\
02 & 025 &   13.43 & 0.356 & 08 & 01 & 001 &   6.97 &  17.28 & 4.179 & 1.002\\
03 & 038 &   33.48 & 0.484 & 22 & 07 & 023 &   5.22 &   8.05 & 0.394 & 1.021\\
03 & 040 &   61.26 & 0.526 & 06 & 17 & 013 &   5.30 &   4.24 & 2.055 & 1.027\\
04 & 002 &    8.52 & 0.271 & 45 & 04 & 045 &   6.46 &  11.09 & 0.207 & 1.005\\
04 & 004 &  118.33 & 0.475 & 04 & 16 & 023 &   8.36 &   7.39 & 2.092 & 1.039\\
04 & 006 &   62.48 & 0.309 & 20 & 01 & 174 &   7.54 &   6.41 & 0.263 & 0.995\\
04 & 007 &   26.88 & 1.000 & 49 & 06 & 062 &   9.48 &   7.92 & 0.055 & 0.992\\
04 & 010 &   26.01 & 0.393 & 06 & 16 & 028 &   5.42 &   4.64 & 2.079 & 0.994\\
04 & 011 &  189.53 & 0.989 & 33 & 15 & 015 &   8.46 &   6.80 & 0.193 & 1.059\\
04 & 012 &    3.82 & 0.352 & 11 & 01 & 163 &   8.96 &  22.71 & 0.242 & 1.000\\
04 & 013 &   16.22 & 0.867 & 09 & 11 & 076 &   5.37 &   4.30 & 0.367 & 1.013\\
04 & 014 &   11.49 & 0.861 & 15 & 02 & 085 &   6.65 &  16.89 & 0.118 & 1.006\\
04 & 015 &   15.34 & 0.997 & 27 & 01 & 150 &   4.84 &  15.69 & 0.045 & 1.001\\
04 & 016 &   26.02 & 0.975 & 04 & 19 & 019 &   7.75 &   7.50 & 0.856 & 1.052\\
04 & 022 &   37.69 & 0.735 & 22 & 14 & 018 &   5.34 &   7.54 & 0.326 & 1.045\\
04 & 023 &   23.30 & 0.722 & 23 & 19 & 010 &   3.56 &   6.20 & 0.394 & 1.031\\
04 & 030 &   61.93 & 0.855 & 08 & 01 & 241 &   4.89 &  17.15 & 0.154 & 1.010\\
04 & 031 &   42.21 & 0.593 & 11 & 15 & 017 &   4.47 &  11.85 & 0.692 & 1.041\\
04 & 032 &  102.10 & 0.281 & 13 & 19 & 023 &   7.01 &   7.29 & 1.307 & 1.050\\

\hline
\end{tabular}
\label{lda_res_tab}
\end{sidewaystable*}

\subsection{Single Strand Results}\label{rslt_sing}

We have also tested the capability of using a single strand to match the observations. This process was performed with the same method as the multi-stranded results, but the number of strands was fixed to one. We tested using a single triangular heating event and multiple heating events contained with the lambda envelope discussed earlier (see Figure~\ref{heatenv}). Examples of the fits are shown by the red and black lines in Figure~\ref{samp_fit_good}. For the case of a single strand heated once, the $\chi^2$ values in Table~\ref{one_res_tab} show that while in a few cases, the single strand heated once can properly match the observations, as a whole, the dynamics of the energization are too complex to be rectified with a single heating event in a 0D strand. The $\chi^2$ and integral ratio for the singly heated fits are consistently larger than for other cases.

\begin{sidewaystable*}[]
\caption{Best fit results from the using a single strand heated only once.}
\begin{tabular}{c|c|c|c|c|c|c|c|c}
\hline\hline

 Data & ARTB & $\chi^2$ & Peak Heating & Heat Pulse & Half-Length & Half-Length & Strand & Integral\\
 Set & Number & (DN) & (ergs/cm$^{3}$/s) & Width (s) & Observed (Mm) & Result (Mm) & Radius (Mm) & Ratio\\
\hline
01 & 000 &  874.65 & 0.994 & 49 &   8.57 &   7.43 & 0.335 & 1.410\\
01 & 001 &   12.41 & 0.854 & 34 &   6.78 &   5.94 & 0.338 & 0.995\\
02 & 001 &  135.70 & 0.999 & 49 &   4.91 &   5.08 & 0.204 & 1.140\\
02 & 002 &  359.19 & 0.997 & 49 &   6.18 &   8.38 & 0.234 & 1.305\\
02 & 004 &  210.79 & 0.999 & 49 &   2.97 &   5.51 & 0.218 & 1.064\\
02 & 010 &   72.13 & 0.995 & 48 &   5.37 &  18.33 & 0.220 & 1.076\\
02 & 014 &  504.65 & 0.997 & 49 &   4.24 &   5.41 & 0.339 & 1.038\\
02 & 015 &  107.94 & 0.996 & 49 &   8.58 &  11.52 & 0.141 & 1.034\\
02 & 017 &   87.57 & 0.927 & 02 &   4.03 &  20.12 & 6.557 & 0.987\\
02 & 018 &   87.69 & 0.612 & 03 &   5.37 &  13.69 & 3.766 & 1.007\\
02 & 020 &  111.64 & 0.980 & 49 &   7.83 &   8.90 & 0.225 & 1.000\\
02 & 021 &   67.27 & 0.947 & 31 &  12.20 &  25.79 & 0.354 & 1.043\\
02 & 022 &   85.92 & 0.795 & 46 &   6.49 &  23.84 & 0.270 & 1.118\\
02 & 023 &   57.64 & 0.998 & 49 &   6.68 &   8.76 & 0.200 & 1.059\\
02 & 024 &   27.41 & 0.999 & 49 &   3.59 &   6.03 & 0.201 & 1.012\\
02 & 025 &   13.16 & 0.494 & 05 &   6.97 &  16.72 & 4.008 & 0.993\\
03 & 038 &   72.17 & 0.969 & 45 &   5.22 &   9.83 & 0.256 & 1.036\\
03 & 040 &  301.73 & 0.998 & 49 &   5.30 &   8.48 & 0.253 & 1.124\\
04 & 002 &   17.54 & 0.997 & 49 &   6.46 &  11.71 & 0.207 & 1.030\\
04 & 004 &  399.09 & 0.877 & 43 &   8.36 &  22.56 & 0.280 & 1.196\\
04 & 006 &  103.78 & 1.000 & 49 &   7.54 &  11.42 & 0.224 & 1.001\\
04 & 007 &  215.38 & 0.993 & 49 &   9.48 &  14.35 & 0.227 & 1.063\\
04 & 010 &  388.82 & 0.999 & 49 &   5.42 &  13.88 & 0.251 & 1.100\\
04 & 011 &  613.44 & 0.999 & 49 &   8.46 &  11.70 & 0.287 & 1.159\\
04 & 012 &    7.58 & 0.305 & 31 &   8.96 &  19.55 & 0.776 & 1.014\\
04 & 013 &   67.22 & 0.964 & 49 &   5.37 &  19.54 & 0.218 & 1.075\\
04 & 014 &   19.01 & 0.801 & 22 &   6.65 &  15.55 & 0.497 & 1.002\\
04 & 015 &   20.69 & 0.423 & 45 &   4.84 &  14.89 & 0.450 & 0.987\\
04 & 016 &   83.55 & 0.922 & 40 &   7.75 &  23.66 & 0.216 & 1.144\\
04 & 022 &   83.15 & 0.976 & 49 &   5.34 &   8.99 & 0.248 & 0.998\\
04 & 023 &   79.27 & 0.991 & 49 &   3.56 &   8.87 & 0.246 & 1.055\\
04 & 030 &  105.19 & 0.814 & 19 &   4.89 &  18.64 & 0.571 & 1.036\\
04 & 031 &   81.15 & 0.999 & 49 &   4.47 &  15.61 & 0.244 & 1.087\\
04 & 032 &  161.86 & 0.745 & 12 &   7.01 &  11.47 & 1.283 & 1.019\\

\hline
\end{tabular}
\label{one_res_tab}
\end{sidewaystable*}

In addition to using a single strand with a single impulsive heating event, we have also matched the observations to a single strand with multiple heating events. This is done to test the usefulness of having independently cooling structures. Since the sinusoidal envelope was consistently less capable, we used the lambda heating envelope and constrained all the heating to occur within a single strand. These results (Table~\ref{mon_res_tab}) are quite interesting: while a single strand heated just once was unable to mimic the observations, a single strand heated multiple times performed similarly (and sometimes better) than multiple independent strands. This promotes the tendency of the multi-stranded cases to favor fewer strands. The best-fit single strand cases had a median of 24 heating events.

\begin{figure*}[]
\includegraphics[width=1.\linewidth]{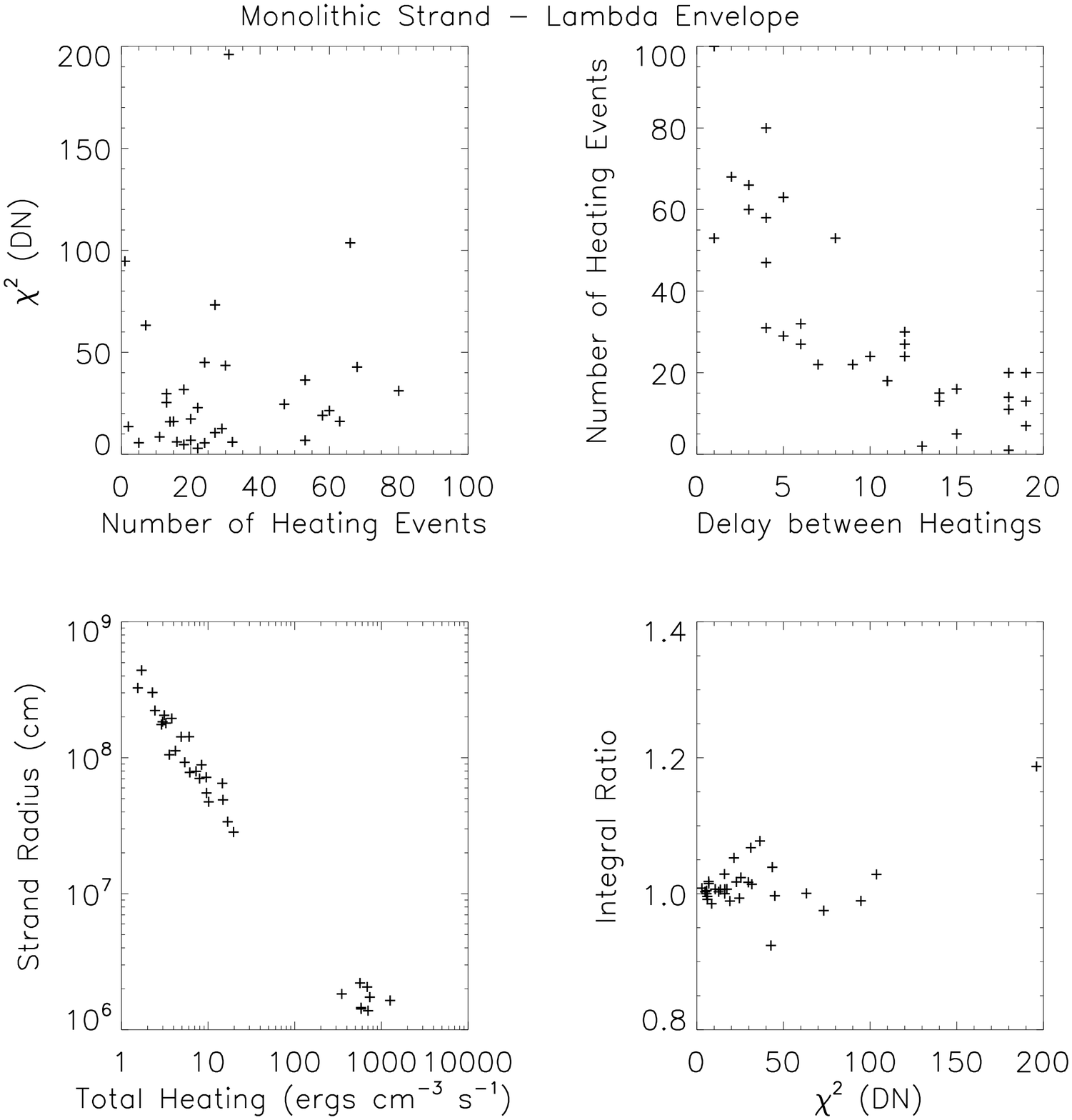}
\caption{Scatter plots for best single strand fits with a lambda heating envelope, the layout is the same as Figure~\ref{res_scat_sin}. In this case, the relation between $\Delta t$ and $n$ has a Spearman coefficient of -0.8819 with a chance probability of less than 0.01\%. As expected, the strand radius is strongly correlated to the total heating input into the strand.}
\label{res_scat}
\end{figure*}

\begin{sidewaystable*}[]
\caption{Results from the using a single strand with multiple heating events (lambda envelope).}
\begin{tabular}{c|c|c|c|c|c|c|c|c|c|c}
\hline\hline
 Data & ARTB & $\chi^2$ & Peak Heating & Heat Pulse & Event & Number of & Half-Length & Half-Length & Strand & Integral\\
 Set & Number & (DN) & (ergs/cm$^{3}$/s) & Width (s) & Delay (s) & Heatings &  Observed (Mm) & Result (Mm) & Radius (Mm) & Ratio\\
\hline
01 & 000 &  776.24 & 0.999 & 49 & 01 & 100 &   8.57 &   8.18 & 0.016 & 1.381\\
01 & 001 &    5.68 & 0.299 & 12 & 15 & 005 &   6.78 &   6.46 & 0.926 & 1.005\\
02 & 001 &   36.41 & 0.983 & 43 & 01 & 053 &   4.91 &   4.68 & 0.014 & 1.078\\
02 & 002 &  196.02 & 0.200 & 05 & 04 & 031 &   6.18 &   5.89 & 0.718 & 1.187\\
02 & 004 &   42.78 & 0.969 & 41 & 02 & 068 &   2.97 &   2.83 & 0.014 & 0.924\\
02 & 010 &    5.63 & 0.237 & 02 & 12 & 024 &   5.37 &   9.89 & 1.843 & 1.000\\
02 & 014 &   63.25 & 0.609 & 12 & 19 & 007 &   4.24 &   4.21 & 0.649 & 1.001\\
02 & 015 &   31.16 & 0.958 & 18 & 04 & 080 &   8.58 &  11.99 & 0.018 & 1.068\\
02 & 017 &   94.66 & 0.359 & 05 & 18 & 001 &   4.03 &  20.12 & 6.766 & 0.990\\
02 & 018 &   25.40 & 0.214 & 02 & 19 & 013 &   5.37 &  15.42 & 3.264 & 1.024\\
02 & 020 &   31.75 & 0.200 & 03 & 11 & 018 &   7.83 &   9.91 & 1.946 & 1.014\\
02 & 021 &   17.33 & 0.200 & 02 & 18 & 020 &  12.20 &  28.16 & 2.226 & 1.007\\
02 & 022 &    6.91 & 0.200 & 04 & 19 & 020 &   6.49 &  13.92 & 1.127 & 1.015\\
02 & 023 &   21.43 & 0.200 & 02 & 03 & 060 &   6.68 &   9.24 & 0.781 & 1.053\\
02 & 024 &    6.11 & 0.874 & 04 & 15 & 016 &   3.59 &   3.44 & 0.490 & 0.992\\
02 & 025 &   13.62 & 0.601 & 03 & 13 & 002 &   6.97 &  17.60 & 4.397 & 1.006\\
03 & 038 &   22.84 & 0.278 & 02 & 07 & 022 &   5.22 &   9.23 & 1.799 & 1.017\\
03 & 040 &   29.74 & 0.200 & 11 & 14 & 013 &   5.30 &   5.95 & 0.888 & 1.017\\
04 & 002 &    4.77 & 0.202 & 09 & 11 & 018 &   6.46 &  11.82 & 0.551 & 1.003\\
04 & 004 &   16.15 & 0.205 & 06 & 05 & 063 &   8.36 &   8.64 & 0.284 & 1.007\\
04 & 006 &   73.25 & 0.200 & 11 & 06 & 027 &   7.54 &  11.08 & 0.338 & 0.975\\
04 & 007 &   19.09 & 0.997 & 49 & 04 & 058 &   9.48 &  11.96 & 0.017 & 0.989\\
04 & 010 &   24.55 & 0.984 & 47 & 04 & 047 &   5.42 &  10.98 & 0.022 & 0.994\\
04 & 011 &  103.69 & 0.975 & 42 & 03 & 066 &   8.46 &   8.06 & 0.021 & 1.029\\
04 & 012 &    2.94 & 0.306 & 02 & 09 & 022 &   8.96 &  20.67 & 1.053 & 1.008\\
04 & 013 &   10.62 & 0.241 & 06 & 12 & 027 &   5.37 &  15.72 & 0.475 & 1.007\\
04 & 014 &    6.03 & 0.242 & 03 & 06 & 032 &   6.65 &  16.16 & 0.704 & 0.996\\
04 & 015 &    8.54 & 0.606 & 03 & 18 & 011 &   4.84 &  15.93 & 0.794 & 0.985\\
04 & 016 &    6.85 & 0.899 & 48 & 08 & 053 &   7.75 &  10.20 & 0.015 & 1.018\\
04 & 022 &   16.09 & 0.591 & 02 & 14 & 015 &   5.34 &   7.07 & 1.426 & 1.000\\
04 & 023 &   12.60 & 0.202 & 03 & 05 & 029 &   3.56 &   5.69 & 1.428 & 1.003\\
04 & 030 &   45.02 & 0.200 & 02 & 10 & 024 &   4.89 &  18.84 & 1.756 & 0.997\\
04 & 031 &   15.98 & 0.263 & 02 & 18 & 014 &   4.47 &  13.13 & 3.020 & 1.029\\
04 & 032 &   43.55 & 0.200 & 02 & 12 & 030 &   7.01 &  12.92 & 2.054 & 1.039\\

\hline
\end{tabular}
\label{mon_res_tab}
\end{sidewaystable*}

\section{Conclusion}\label{conc}
We were able to successfully forward model 34 ARTBs observed with XRT as multi-stranded and single-stranded structures using the 0D model EBTEL. We varied the heating rate (spatially via multiple strands and temporally with multiple strands and heating events) to study the energy deposition in these small brightenings. An overview of the results is found in Table~\ref{artb_tab}. It was found that for most events, multiple heating events were required. For a multi-stranded system heated with a sinusoidal envelope, a median of 31 strands were required to reproduce the observations. For the lambda envelope, the multi-stranded cases used a median of 23 strands, and the single strand cases used a median of 24 heating events to reproduce the observed ARTB evolution. 

 Additionally, we were able to show that a symmetric (sinusoidal) envelope to the heating was not as capable as a more impulsive envelope. This supports the concept of a ``triggering'' mechanism to these events, wherein something causes a quick burst of heating to these structures, and then quickly fades, as opposed to a more uniform build up and decline to the heating. This tendency towards asymmetric (in time) energization appears to be consistent with the abrupt deposition of energy one might expect from a reconnection-related micro-flaring scenario.  On the other hand, wave-heating scenarios, wherein energy is deposited in the coronal loop via a gradually developing resonance within the loop, might be expected to ``switch on'' less abruptly, and thus may favor the sinusoidal heating profile.  The frequency with which the lambda or sinusoidal heating profile achieves a better fit to the observed ARTBs, which may yield inferences regarding the relative importance of reconnection- versus wave-heating, will be addressed by a future study involving a much larger catalog of ARTBs.  The present proof-of-concept study serves to introduce the methodology and demonstrate its abilities and limitations in modeling the ARTBs.

One thing we were not able to discern is whether these events require multiple structures in the cross-section perpendicular to the loop axis or if the observed evolution is caused by multiple heating events within a single structure. If considering these events to be the result of braided loops heated by reconnecting braids (causing the braids to re-configure in a less twisted fashion), it is likely that each strand reconnects multiple times before the system stabilizes in a less twisted state, and thus multiple heating events for each strand should be expected. An important distinction between the multi- and single-stranded cases is whether a strand is allowed to cool completely after individual heating events. The multi-stranded cases represent the extreme end of low frequency heating, where the strand is allowed to cool completely before being heated again. The single-stranded cases allow high frequency heating (minimal cooling before being heated again) as well as low frequency heating. Our single-strand results favor high frequency heating, but since we cannot yet find a preference for the multi-strand or single-strand results, we cannot yet reliably determine the frequency of heating for these events, similar to the results found in \cite{Guennou2013}. 

It would be very useful to understand better the number and frequency of heating events each strand experiences. In the case of multi-stranded loops, however, adding the ability for each strand to experience multiple heating events runs the risk of over-constraining the system, i.e. it would require more parameters than data points for each observation. It is difficult to properly constrain the model using multiple strands heated multiple times without over constraining the data: the model would have to constrain both the number of heating events per strand and the envelope between them in addition to the envelope between individual strands used here. It is our hope that this and future studies will help to constrain the models such that it will be possible to adequately test the effects of multiple strands heated multiple times. Additionally, since our results cannot yet distinguish between the multi-stranded singly heated and single-strand with multiple heating events, it is difficult to justify that this method as currently implemented would be able to distinguish the models shown here with the case of multiple strands heated multiple times. The next step in approaching this level of complexity will be to find many more ARTBs in the XRT data catalog and model them to more rigorously constrain the requirements on the number of strands (and heating events) necessary.

\acknowledgments
This work was partially supported by NASA under contract NNM07AB07C with the Smithsonian Astrophysical Observatory.

%\facility{XRT}
\bibliographystyle{apj}       % APS-like style for physics
\bibliography{kobelski}

\newcommand{\noop}[1]{}
\begin{thebibliography}{}
\expandafter\ifx\csname natexlab\endcsname\relax\def\natexlab#1{#1}\fi

\bibitem[{{Berghmans} \& {Clette}(1999)}]{BerghmansClette}
{Berghmans}, D., \& {Clette}, F. 1999, \solphys, 186, 207

\bibitem[{{Berghmans} {et~al.}(2001){Berghmans}, {McKenzie}, \&
  {Clette}}]{BerghmansMcKenzieClette}
{Berghmans}, D., {McKenzie}, D., \& {Clette}, F. 2001, \aap, 369, 291

\bibitem[{{Brooks} {et~al.}(2012){Brooks}, {Warren}, \&
  {Ugarte-Urra}}]{Brooks2012}
{Brooks}, D.~H., {Warren}, H.~P., \& {Ugarte-Urra}, I. 2012, \apjl, 755, L33

\bibitem[{{Cargill}(1994)}]{Cargill1994}
{Cargill}, P.~J. 1994, \apj, 422, 381

\bibitem[{{Cargill} {et~al.}(2012{\natexlab{a}}){Cargill}, {Bradshaw}, \&
  {Klimchuk}}]{EBTEL2}
{Cargill}, P.~J., {Bradshaw}, S.~J., \& {Klimchuk}, J.~A. 2012{\natexlab{a}},
  \apj, 752, 161

\bibitem[{{Cargill} {et~al.}(2012{\natexlab{b}}){Cargill}, {Bradshaw}, \&
  {Klimchuk}}]{EBTEL3}
---. 2012{\natexlab{b}}, \apj, 758, 5

\bibitem[{{Cargill} {et~al.}(1995){Cargill}, {Mariska}, \&
  {Antiochos}}]{CargillModel1995}
{Cargill}, P.~J., {Mariska}, J.~T., \& {Antiochos}, S.~K. 1995, \apj, 439, 1034

\bibitem[{{Charbonneau}(1995)}]{pikaia}
{Charbonneau}, P. 1995, \apjs, 101, 309

\bibitem[{{Cirtain} {et~al.}(2013){Cirtain}, {Golub}, {Winebarger}, {de
  Pontieu}, {Kobayashi}, {Moore}, {Walsh}, {Korreck}, {Weber}, {McCauley},
  {Title}, {Kuzin}, \& {Deforest}}]{CirtainHiC2013}
{Cirtain}, J.~W., {Golub}, L., {Winebarger}, A.~R., {et~al.} 2013, \nat, 493,
  501

\bibitem[{{Culhane} {et~al.}(2007){Culhane}, {Harra}, {James}, {Al-Janabi},
  {Bradley}, {Chaudry}, {Rees}, {Tandy}, {Thomas}, {Whillock}, {Winter},
  {Doschek}, {Korendyke}, {Brown}, {Myers}, {Mariska}, {Seely}, {Lang}, {Kent},
  {Shaughnessy}, {Young}, {Simnett}, {Castelli}, {Mahmoud}, {Mapson-Menard},
  {Probyn}, {Thomas}, {Davila}, {Dere}, {Windt}, {Shea}, {Hagood}, {Moye},
  {Hara}, {Watanabe}, {Matsuzaki}, {Kosugi}, {Hansteen}, \&
  {Wikstol}}]{EISInstrument2007}
{Culhane}, J.~L., {Harra}, L.~K., {James}, A.~M., {et~al.} 2007, \solphys, 243,
  19

\bibitem[{{Deforest} \& {Gurman}(1998)}]{DeForest1998}
{Deforest}, C.~E., \& {Gurman}, J.~B. 1998, \apjl, 501, L217

\bibitem[{{Delaboudini{\`e}re} {et~al.}(1995){Delaboudini{\`e}re}, {Artzner},
  {Brunaud}, {Gabriel}, {Hochedez}, {Millier}, {Song}, {Au}, {Dere}, {Howard},
  {Kreplin}, {Michels}, {Moses}, {Defise}, {Jamar}, {Rochus}, {Chauvineau},
  {Marioge}, {Catura}, {Lemen}, {Shing}, {Stern}, {Gurman}, {Neupert},
  {Maucherat}, {Clette}, {Cugnon}, \& {van Dessel}}]{EIT1995}
{Delaboudini{\`e}re}, J.-P., {Artzner}, G.~E., {Brunaud}, J., {et~al.} 1995,
  \solphys, 162, 291

\bibitem[{{Golub} {et~al.}(2007){Golub}, {Deluca}, {Austin}, {Bookbinder},
  {Caldwell}, {Cheimets}, {Cirtain}, {Cosmo}, {Reid}, {Sette}, {Weber},
  {Sakao}, {Kano}, {Shibasaki}, {Hara}, {Tsuneta}, {Kumagai}, {Tamura},
  {Shimojo}, {McCracken}, {Carpenter}, {Haight}, {Siler}, {Wright}, {Tucker},
  {Rutledge}, {Barbera}, {Peres}, \& {Varisco}}]{GolubXRT}
{Golub}, L., {Deluca}, E., {Austin}, G., {et~al.} 2007, \solphys, 243, 63

\bibitem[{{Guennou} {et~al.}(2013){Guennou}, {Auch{\`e}re}, {Klimchuk},
  {Bocchialini}, \& {Parenti}}]{Guennou2013}
{Guennou}, C., {Auch{\`e}re}, F., {Klimchuk}, J.~A., {Bocchialini}, K., \&
  {Parenti}, S. 2013, \apj, 774, 31

\bibitem[{{Kano} {et~al.}(2008){Kano}, {Sakao}, {Hara}, {Tsuneta}, {Matsuzaki},
  {Kumagai}, {Shimojo}, {Minesugi}, {Shibasaki}, {Deluca}, {Golub},
  {Bookbinder}, {Caldwell}, {Cheimets}, {Cirtain}, {Dennis}, {Kent}, \&
  {Weber}}]{Kano2008}
{Kano}, R., {Sakao}, T., {Hara}, H., {et~al.} 2008, \solphys, 249, 263

\bibitem[{{Klimchuk} {et~al.}(1992){Klimchuk}, {Lemen}, {Feldman}, {Tsuneta},
  \& {Uchida}}]{Klimchuketal1992}
{Klimchuk}, J.~A., {Lemen}, J.~R., {Feldman}, U., {Tsuneta}, S., \& {Uchida},
  Y. 1992, \pasj, 44, L181

\bibitem[{{Klimchuk} {et~al.}(2008){Klimchuk}, {Patsourakos}, \&
  {Cargill}}]{EBTEL1}
{Klimchuk}, J.~A., {Patsourakos}, S., \& {Cargill}, P.~J. 2008, \apj, 682, 1351

\bibitem[{{Kobelski} {et~al.}(\noop{3001}in press){Kobelski}, {Saar}, {Weber},
  {McKenzie}, \& {Reeves}}]{xrtprep}
{Kobelski}, A., {Saar}, S., {Weber}, M., {McKenzie}, D., \& {Reeves}, K.
  \noop{3001}in press, \solphys

\bibitem[{{Kosugi} {et~al.}(2007){Kosugi}, {Matsuzaki}, {Sakao}, {Shimizu},
  {Sone}, {Tachikawa}, {Hashimoto}, {Minesugi}, {Ohnishi}, {Yamada}, {Tsuneta},
  {Hara}, {Ichimoto}, {Suematsu}, {Shimojo}, {Watanabe}, {Shimada}, {Davis},
  {Hill}, {Owens}, {Title}, {Culhane}, {Harra}, {Doschek}, \&
  {Golub}}]{Kosugi2007}
{Kosugi}, T., {Matsuzaki}, K., {Sakao}, T., {et~al.} 2007, \solphys, 243, 3

\bibitem[{{Lemen} {et~al.}(2012){Lemen}, {Title}, {Akin}, {Boerner}, {Chou},
  {Drake}, {Duncan}, {Edwards}, {Friedlaender}, {Heyman}, {Hurlburt}, {Katz},
  {Kushner}, {Levay}, {Lindgren}, {Mathur}, {McFeaters}, {Mitchell}, {Rehse},
  {Schrijver}, {Springer}, {Stern}, {Tarbell}, {Wuelser}, {Wolfson}, {Yanari},
  {Bookbinder}, {Cheimets}, {Caldwell}, {Deluca}, {Gates}, {Golub}, {Park},
  {Podgorski}, {Bush}, {Scherrer}, {Gummin}, {Smith}, {Auker}, {Jerram},
  {Pool}, {Soufli}, {Windt}, {Beardsley}, {Clapp}, {Lang}, \&
  {Waltham}}]{AIAInstrument2012}
{Lemen}, J.~R., {Title}, A.~M., {Akin}, D.~J., {et~al.} 2012, \solphys, 275, 17

\bibitem[{{Narukage} {et~al.}(2011){Narukage}, {Sakao}, {Kano}, {Hara},
  {Shimojo}, {Bando}, {Urayama}, {Deluca}, {Golub}, {Weber}, {Grigis},
  {Cirtain}, \& {Tsuneta}}]{Narukage2011}
{Narukage}, N., {Sakao}, T., {Kano}, R., {et~al.} 2011, \solphys, 269, 169

\bibitem[{{Ofman} {et~al.}(1995){Ofman}, {Davila}, \&
  {Steinolfson}}]{Ofmanetal1995}
{Ofman}, L., {Davila}, J.~M., \& {Steinolfson}, R.~S. 1995, \apj, 444, 471

\bibitem[{{Parker}(1988)}]{Parker1988}
{Parker}, E.~N. 1988, \apj, 330, 474

\bibitem[{{Pesnell} {et~al.}(2012){Pesnell}, {Thompson}, \&
  {Chamberlin}}]{SDOInstrument2012}
{Pesnell}, W.~D., {Thompson}, B.~J., \& {Chamberlin}, P.~C. 2012, \solphys,
  275, 3

\bibitem[{{Reeves} \& {Warren}(2002)}]{ReevesWarren}
{Reeves}, K.~K., \& {Warren}, H.~P. 2002, \apj, 578, 590

\bibitem[{{Sarkar} \& {Walsh}(2008)}]{SarkarWalsh2008}
{Sarkar}, A., \& {Walsh}, R.~W. 2008, \apj, 683, 516

\bibitem[{{Schmelz} {et~al.}(2007){Schmelz}, {Roames}, \&
  {Nasraoui}}]{Schmelzetal2007}
{Schmelz}, J.~T., {Roames}, J.~K., \& {Nasraoui}, K. 2007, Advances in Space
  Research, 39, 1497

\bibitem[{{Seaton} {et~al.}(2001){Seaton}, {Winebarger}, {DeLuca}, {Golub},
  {Reeves}, \& {Gallagher}}]{Seaton2001}
{Seaton}, D.~B., {Winebarger}, A.~R., {DeLuca}, E.~E., {et~al.} 2001, \apjl,
  563, L173

\bibitem[{{Shimizu}(1995)}]{Shimizu1995}
{Shimizu}, T. 1995, \pasj, 47, 251

\bibitem[{{Shimizu} {et~al.}(1992){Shimizu}, {Tsuneta}, {Acton}, {Lemen}, \&
  {Uchida}}]{Shimizuetal1992}
{Shimizu}, T., {Tsuneta}, S., {Acton}, L.~W., {Lemen}, J.~R., \& {Uchida}, Y.
  1992, \pasj, 44, L147

\bibitem[{{Tsuneta} {et~al.}(1991){Tsuneta}, {Acton}, {Bruner}, {Lemen},
  {Brown}, {Caravalho}, {Catura}, {Freeland}, {Jurcevich}, {Morrison},
  {Ogawara}, {Hirayama}, \& {Owens}}]{SXT1991}
{Tsuneta}, S., {Acton}, L., {Bruner}, M., {et~al.} 1991, \solphys, 136, 37

\bibitem[{{Ugarte-Urra} \& {Warren}(2013)}]{Ignacio2013}
{Ugarte-Urra}, I., \& {Warren}, H.~P. 2013, ArXiv e-prints, arXiv:1311.6346

\bibitem[{{Warren}(2006)}]{Warren2006}
{Warren}, H.~P. 2006, \apj, 637, 522

\bibitem[{{Yoshida} {et~al.}(1995){Yoshida}, {Tsuneta}, {Golub}, {Strong}, \&
  {Ogawara}}]{Yoshidaetal1995}
{Yoshida}, T., {Tsuneta}, S., {Golub}, L., {Strong}, K., \& {Ogawara}, Y. 1995,
  \pasj, 47, L15

\end{thebibliography}

\clearpage

\end{document}